# Linear Enhancement of Spin-Orbit Torques and Absence of Bulk Rashba-Type Spin Splitting in Perpendicularly Magnetized [Pt/Co/W]$_n$ Superlattices

Zhihao Yan[1,2], Zhengxiao Li[1,2], Lujun Zhu[3], Xin Lin[1,2], Lijun Zhu[1,2]**

1. *Institute of Semiconductors, Chinese Academy of Sciences, Beijing 100083, China*
2. *Center of Materials Science and Optoelectronics Engineering, University of Chinese Academy of Sciences, Beijing 100049, China*
3. *College of Physics and Information Technology, Shaanxi Normal University, Xi'an 710062, China*

*ljzhu@semi.ac.cn

**Abstract:** The development of magnetic heterostructures with strong spin-orbit torques (SOTs), low impedance, strong perpendicular magnetic anisotropy (PMA), and good integration compatibility at the same time is central for high-performance spintronic memory and computing applications. Here, we report the development of the symmetry-broken spin-orbit superlattice [Pt/Co/W]$_n$ that can be sputtered-deposited on commercial oxidized silicon substrates and have giant SOTs, strong uniaxial PMA of ≈9.2 Merg/cm$^3$. The dampinglike and fieldlike SOTs of the [Pt/Co/W]$_n$ superlattices exhibit a linear increase with the repeat number *n* and reach the giant values of 225% and -33% (two orders of magnitude greater than that in clean-limit Pt) at *n* = 12, respectively. The dampinglike SOT is also of the opposite sign and much greater in magnitude than the fieldlike SOT, regardless of the number of *n*. These results clarify that the spin current that generates SOTs in the [Pt/Co/W]$_n$ superlattices arises predominantly from the spin Hall effect rather than bulk Rashba-type spin splitting, providing a unified understanding of the SOTs in the superlattices. We also demonstrate deterministic switching in thicker-than-50-nm PMA [Pt/Co/W]$_{12}$ superlattices at a low current density. This work establishes the [Pt/Co/W]$_n$ superlattice as a compelling material candidate for ultra-fast, low-power, long-retention nonvolatile spintronic memory and computing technologies.

## I. Introduction

It has been a core topic of spintronics to develop magnetic heterostructures that have high spin-orbit torques (SOTs), low impedance, strong perpendicular magnetic anisotropy (PMA), and good integration compatibility at the same time, for ultra-fast, low-power, long-retention nonvolatile spintronic memory and computing technologies.[1-7] However, the SOT efficiencies of the conventional spin Hall metal/ferromagnet bilayers are usually low, e.g., the dampinglike SOT efficiency ($\xi_{DL}^j$) is only 0.02 for Pt/ferromagnet bilayers with a clean-limit Pt (resistivity ≈ 7 μΩ cm[8]) and 0.05-0.1 for that with more dirty Pt.[9-14]

We have previously proposed [15] and demonstrated that a symmetry-broken spin-orbit superlattice, which stacks identical trilayer of the spin-current generator/magnet/symmetry-breaking layer with the repeat number (*n*), increases the dampinglike and fieldlike SOT efficiencies, $\xi_{DL}^j$ and $\xi_{FL}^j$, by *n* times and towards infinity (e.g., $\xi_{DL}^j$ = 2.1 in *in-plane anisotropy* [Pt$_{0.75}$Cu$_{0.25}$/Co/Ta]$_n$ superlattices with *n* = 16)[16]. Such symmetry-broken spin-orbit superlattices provide a universal material scheme that has enhanced SOTs, PMA, and conductance at the same time. The linear increase of the dampinglike and fieldlike SOTs with *n* in the [Pt$_{0.75}$Cu$_{0.25}$/Co/Ta]$_n$ superlattices and the absence of SOTs in the symmetry-perservered [Pt/Co]$_n$ superlattices[16] indicate that the spin Hall effect (SHE) is the only source of the spin current that generates SOTs in the superlattices, while the stacking by itself does not generate any observable spin currents via the individual interfaces or any collective effect of the superlattices.

So far, the linear enhancement of the SOTs, however, has not been achieved in any spin-orbit superlattices with PMA. This is mainly because the demonstration of the effects would require good identity of the repeats across the large thickness of the superlattices, which is challenging for most material systems due to possible degradation of the layer smoothness and PMA induced, e.g., by the strain-relaxation, during the stacking of superlattices. For instance, the [Pt/Co/MgO]$_n$ superlattice degrades substantially at large repeat numbers in smoothness, resistivity, and saturation magnetization, [15] leading to much slower SOT enhancement than the expected linear-in-*n* scaling. Stimulatingly, a recent study by Ham *et al.*[17] suggests that stacking Pt/Co/W trilayers could form [Pt/Co/W]$_n$ superlattices with reasonable layer smoothness. However, that work claimed a non-monotonic variation of the SOTs with *n*, low upper-limit SOT efficiencies ($\xi_{\text{DL}}^j \approx 0.4$ and $\xi_{\text{FL}}^j \approx 0.7$), and a dominant SOT source of stacking-collectively-induced "bulk Rashba-type spin splitting" in their [Pt/Co/W]$_n$ superlattices disobeying the macrospin model [17]. These strong disagreements suggest an urgent need to verify the exact *n* dependence, the upper limit, and the physics origin of the SOTs of the perpendicularly magnetized [Pt/Co/W]$_n$ for a unified understanding of various superlattices.

In this work, we report the development, the SOT mechanism, and the practical impacts of [Pt/Co/W]$_n$ superlattices with strong PMA and rigid macrospin behaviors at low magnetic fields. We demonstrate that both dampinglike and fieldlike SOTs of the [Pt/Co/W]$_n$ superlattices exhibit a reasonable linear-in-*n* enhancement towards infinite, with the dampinglike torque being always of the opposite sign and much greater in magnitude than the fieldlike torque. These results clarify that the SOTs in the [Pt/Co/W]$_n$ superlattice arise from the SHE, rather than the bulk Rashba-type spin splitting.

## II. Sample Details

For this study, we fabricated 5 × 60 μm² Hall bars of [Pt (2.5)/Co (0.8)/W (1.5)]$_n$ superlattices (numbers in the brackets are layer thickness in nanometer, the repeat number *n* = 1, 4, 6, 8, 10, and 12, respectively) sputter-deposited on thermally oxidized silicon substrates and protected by MgO(2)/TaO$_x$ bilayers (Fig.1a,b). A thickness of 0.8 nm is chosen for the Co layers as a compromise between the interfacial PMA and spin dephasing [18-20]. The Pt layers are made 2.5 nm thick to generate strong SOTs without making the whole superlattice (*n* ≤12) too thick to etch (over-etching damages the photoresist).

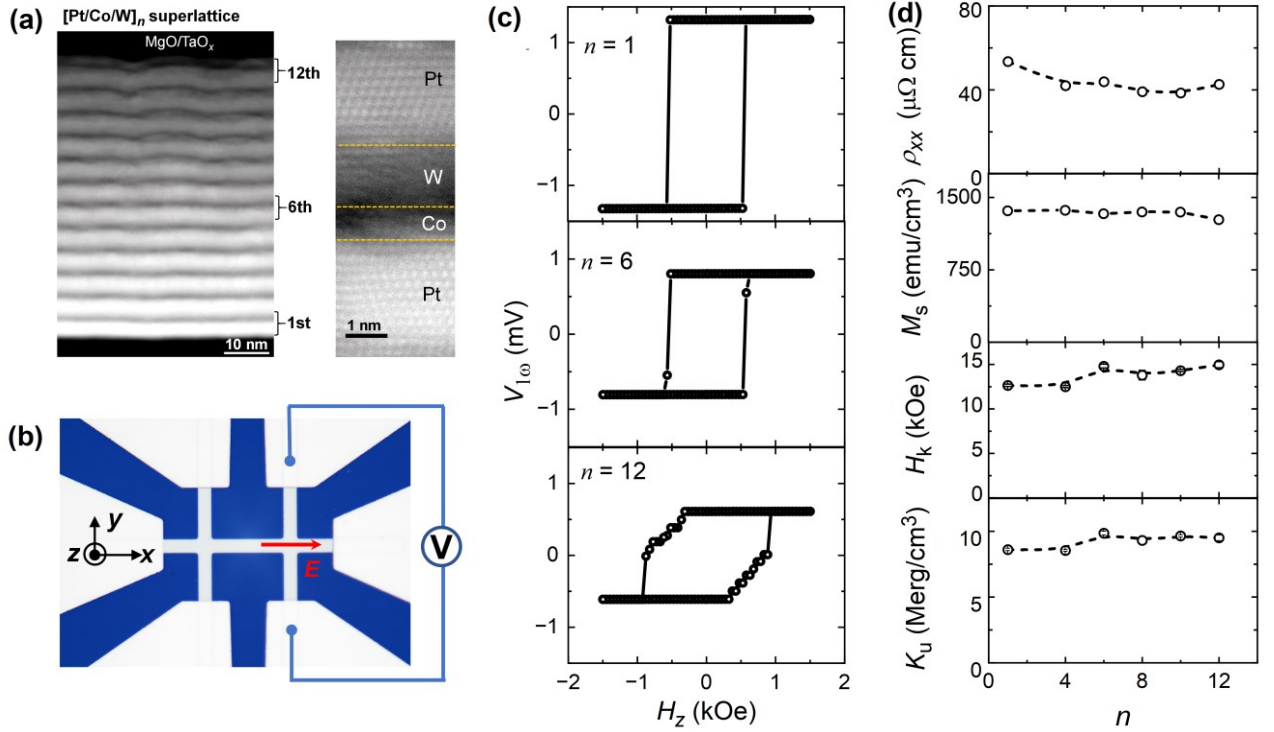

**Figure 1**. (a) High-angle annular dark-field scanning transmission electron microscopy images of the [Pt/Co/W]$_{12}$ superlattice and (b) Optical microscopy image of the Hall bar device. (c) First harmonic Hall voltage vs perpendicular magnetic field, (d) Resistivity, saturation magnetization, and perpendicular magnetic anisotropy field, uniaxial PMA energy of the [Pt/Co/W]$_n$ superlattice with varying repeat numbers $n$.

Cross-sectional high-angle annular dark-field scanning transmission electron microscopy (HAADF STEM) characterizations suggest that the [Pt/Co/W]$_n$ superlattices have very good layer smoothness for $n \leq 6$, but slightly increased roughness for larger $n$ (Fig. 1a). The first harmonic Hall voltage (HHV) hysteresis loops as a function of the perpendicular magnetic field ($H_z$) suggest good PMA (Fig. 1c). In Fig. 1d, we plot the Pt resistivity ($\rho_{xx}$), the saturation magnetization ($M_s$), PMA field ($H_k$), and uniaxial PMA energy ($K_u = M_s H_k / 2$) of the Co layers as a function of $n$ of the superlattices. $\rho_{xx}$ varies from 54 μΩ cm to 39 μΩ cm as estimated from the conductance enhancement of the Ta(1)/[Pt/Co/W]$_n$/MgO(2)/Ta(1.5) stack, relative to $n$ times the conductance of a control stack lacking a Pt layer, i.e., Ta(1)/Co(0.8)/W(1.5)/MgO(1.6)/ Ta(1.5). These superlattices exhibit high $M_s$ of between 1340±36 emu/cm³, strong $H_k$ of 13.8±1.1 kOe, and giant uniaxial PMA energy of 9.2±0.5 Merg/cm³ (much greater than that of the bulk PMA ferrimagnets FeTb [21] and the optimized Ta/CoFeB/MgO[22]).

## III. Linear-in-$n$ enhancement of spin-orbit torques

The SOT efficiencies in the [Pt/Co/W]$_n$ superlattices are carefully characterized using HHV measurements in the low-field macrospin regime by taking into account thermoelectric effects.[15,23] The in-phase first and out-of-phase second HHVs, $V_{1\omega}$ and $V_{2\omega}$, are collected as a function of the swept longitudinal ($H_x$) and transverse ($H_y$) magnetic fields while applying a sinusoidal electric field ($E$) of 30 kV/m onto the Hall bar orientated along the $x$ axis (Fig.1b). For all the superlattices in this study, $V_{1\omega}$ exhibits 100% remanent ratio during the sweeping of $H_z$ (Fig. 1c) and perfect parabolic scaling during the sweeping of $H_x$ and $H_y$, while $V_{2\omega}$ exhibits excellent linear scaling with the in-plane fields (Fig. 2a,b). These characteristics consistently reveal the rigid macrospin behavior [1] of these superlattices in the studied regime of in-plane magnetic fields (e.g., $H_x$, $H_y$ < 1.5 kOe for the superlattice with $n$ = 12). To further ensure the macrospin behavior and avoid the occurrence of any multidomain states during the HHV measurements, we always saturate the superlattices into the +$M_z$ (-$M_z$) sates using a sufficiently large positive (negative) $H_z$ before collecting $V_{1\omega}$ and $V_{2\omega}$ as a function of the in-plane magnetic fields swept *only* in the low field ranges. Note that the rigid fulfillment of the macrospin behaviors in the studied magnetic field regimes is critical for the reliable analysis of the SOTs because otherwise, the HHV technique would not apply at all. The superlattice with $n$ = 12 does show some steplike switching during $H_z$ scanning in Fig. 1c likely due to a sizable distribution of the pinning field, which however, should not affect the macrospin analysis in the regime far away from magnetization switching.

Based on the rigid macrospin behavior of the samples, we determine $H_k$, the dampinglike SOT field ($H_{DL}$), and the fieldlike SOT field ($H_{FL}$) from the parabolic scaling of $V_{1\omega}$ and linear scaling of $V_{2\omega}$ with in-plane magnetic fields, i.e.,

$$V_{1\omega} \approx V_{AHE}(1 - H_{x(y)}^2/2H_k^2), \qquad (1)$$

$$H_{DL} = -2\frac{\partial V_{2\omega}}{\partial H_x} \Big/ \frac{\partial^2 V_{1\omega}}{\partial^2 H_x} - 2H_k V_{ANE,z}/V_{AHE}. \qquad (2)$$

$$H_{FL} = -2\frac{\partial V_{2\omega}}{\partial H_y} \Big/ \frac{\partial^2 V_{1\omega}}{\partial^2 H_y} - H_{Oe}. \qquad (3)$$

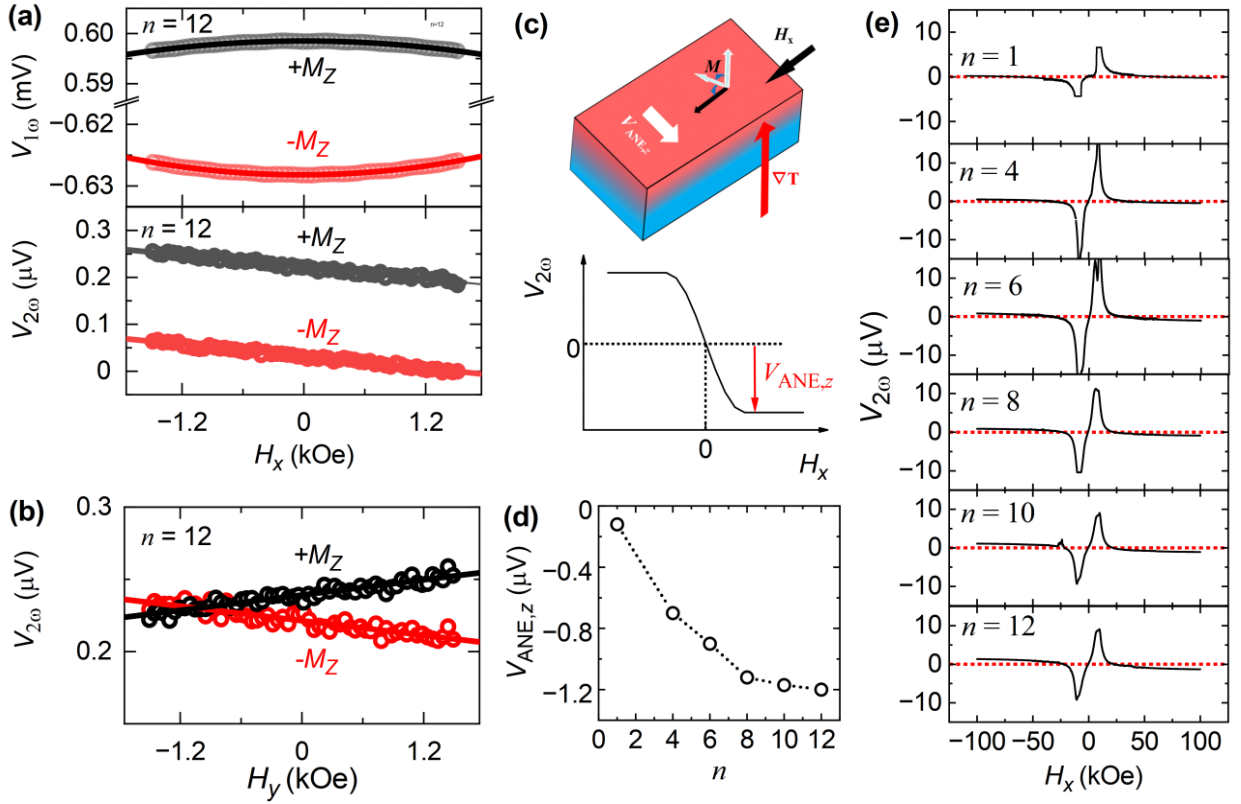

**Figure 2**. First and second harmonic Hall voltages ($V_{1\omega}$, $V_{2\omega}$) of the [Pt(2.5)/Co(0.8)/W (1.5)]$_{12}$ superlattices plotted as a function of (a) the longitudinal magnetic field ($H_x$) and (b) the transverse magnetic field ($H_y$). (c) Schematic of the generation and measurement of the anomalous Nernst voltage of PMA samples. (d) The anomalous Nernst voltage and (e) $V_{2\omega}$ vs in-plane magnetic field for the [Pt/Co/W]$_n$ superlattices ($n$ =1, 4, 6, 8, 10 and 12).

where $V_{AHE}$ is the anomalous Hall voltage that is extracted from the dependence of $V_{1\omega}$ on $H_z$ in Fig. 1c. $H_{Oe}$ is the transverse Oersted field exerted on the Co layers by the in-plane charge current in other layers and can be estimated as $(j_{Pt}d_{Pt} - j_W d_W)/2$ with $j_{Pt(W)}$ and $d_{Pt(W)}$ being the current density and the layer thickness of the Pt (W) layers (see detailed discussions in our previous work [16]). From the fitting of the scaling of $V_{1\omega}$ with $H_x$ and $H_y$ to Eq (1), $H_k$ is determined to 13.8±1.0 kOe for the [Pt/Co/W]$_n$ superlattices with different $n$. $V_{ANE,z}$ is the anomalous Nernst voltage due to the perpendicular thermal gradient and equal to the value of $V_{2\omega}$ when the magnetization is fully aligned along the current direction by $H_x$ of 90 kOe (Fig. 2c). As shown in Fig. 2d,e, $V_{ANE,z}$ becomes increasingly significant as $n$ increases.

With the values of $H_{DL}$ and $H_{FL}$ determined using Eqs. (2) and (3), the dampinglike and field-like SOT efficiencies per electric field, $\xi_{DL}^E$ and $\xi_{FL}^E$, are calculated following[9]

$$\xi_{DL(FL)}^E = \left(\frac{2e}{\hbar}\right)\mu_0 M_s t H_{DL(FL)}/E, \quad (4)$$

where $e$ is the elementary charge, $\hbar$ the reduced Planck's constant, $\mu_0$ the permeability of vacuum, and $t$ the total thickness of the magnetic layer ($t = 0.8n$ nm for the [Pt/Co/W]$_n$ superlattice). As shown in Fig. 3a, both $\xi_{DL}^E$ and $\xi_{FL}^E$ linearly increase in magnitude as $n$ increases, while $\xi_{FL}^E$ is of the opposite sign and 3.5 times smaller in magnitude than $\xi_{DL}^E$.

Since we have only measured negligible SOTs in the control samples Ta (1)/Co (0.8)/W (1.5)/MgO (1.6)/Ta (1.5) and Ta (1)/Co (0.8)/MgO (1.6)/Ta (1.5), the SOTs of the [Pt/Co/W]$_n$ superlattices should be predominantly from the SHE of the Pt layers. This suggests that the W layers in the [Pt/Co/W]$_n$ superlattices are unable to generate detectable spin flow due to its low spin Hall conductivity, insufficient layer thickness, and/or poor crystallinity (Fig. 1a) despite that a thick metastable $\beta$-phased W layer is reported to have a negative spin Hall ratio [24]. This allows to accurately calculate the dampinglike and fieldlike SOT efficiencies per current density, $\xi_{DL}^j$ and $\xi_{DL}^j$, of the [Pt/Co/W]$_n$ superlattices using

$$\xi_{DL(FL)}^j = \rho_{xx}\xi_{DL(FL)}^E, \quad (5)$$

As summarized in Fig. 3a, both $\xi_{DL}^j$ and $\xi_{FL}^j$ linearly increase in magnitude with growing $n$, i.e., $\xi_{DL}^j = 0.173n$ and $\xi_{FL}^j = -0.027n$. The enhancement rate $\xi_{DL}^j/n = 0.173$ for the [Pt (2.5)/Co (0.8)/W (1.5)]$_n$ ($\rho_{xx}$= 39-54 μΩ cm) is slightly greater than that of in-plane magnetized Pt (2)/Co (0.8)/MgO ($\xi_{DL}^j = 0.15$, $\rho_{xx}$= 60 μΩ cm)[16] mainly due to the greater Pt thickness.

**IV. Absence of bulk Rashba-type spin splitting**

The linear-in-$n$ enhancement of the SOTs towards infinity and the fieldlike SOT being of the opposite sign and much smaller in magnitude than the dampinglike SOT agree well with the SHE being the dominant origin of the SOTs in the symmetry-broken spin-orbit superlattices [15,16] and reveal negligible stacking-collectively-induced contribution of SOTs or spin currents.

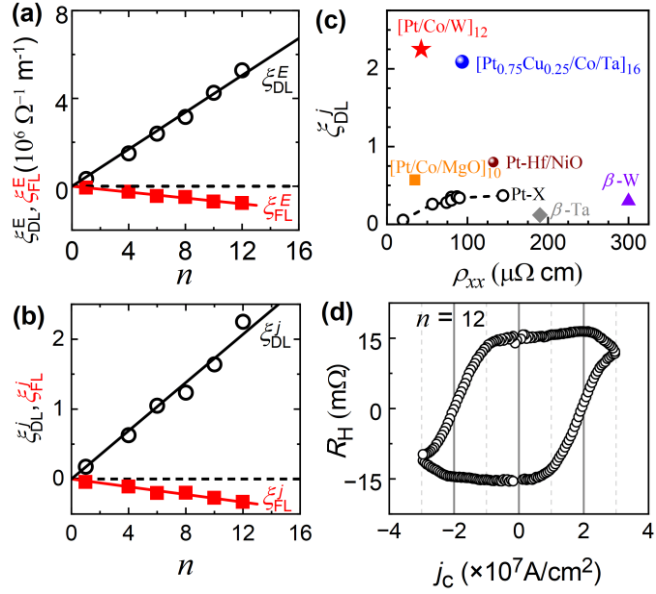

Figure 3. (a) Dampinglike and fieldlike SOT efficiencies per electric field and (b) Damping-like and field-like SOT efficiencies per current density for the of the [Pt(2.5)/Co(0.8)/W(1.5)]$_n$ superlattices, indicating a linear increase of the SOTs with the repeat number $n$. (c) Comparison of and resistivity of strong SOT materials. (d) Hall resistance vs charge current density of the [Pt(2.5)/Co(0.8)/W(1.5)]$_{12}$ ($H_x$ = 3 kOe), showing deterministic switching at an average current density of about 2.0× 10$^7$ A/cm$^2$.

These observations are against the proposal of the bulk Rashba-type spin splitting in previous work [17], the latter was claimed to yield non-monotonic scaling of SOTs with $n$ and a fieldlike torque that is of the same sign but almost twice in magnitude of the dampinglike torque. As we discuss below, the opposite SOT conclusions in the previous work can be attributed to the improper employment of the HHV and the "planar Hall correction" analyses in that work. First, the HHV analysis does not apply to the [Pt/Co/W]$_n$ samples in the previous work[17] since they disobey the macrospin model, the base of the HHV analysis, as revealed by the asymmetry of the $V_{1\omega}$ signals and the remarkable hysteresis of $V_{2\omega}$ signals as a function of the swept longitudinal and transverse magnetic fields ($H_x$, $H_y$, see Fig. 2b,c in Ref. [17]). Furthermore, that analysis employed a large planar Hall correction (the ratio of the planar Hall voltage and the anomalous Hall voltage $V_{PHE}/V_{AHE}$ was up to 0.35). However, as we summarized in Refs. [1] and [14], this correction, since the same year of the proposal (2014) [25], has been known to cause errors by researchers including the developer of this "correction" and should be avoided [13,26-29]. When $V_{PHE}/V_{AHE}$ is very large, this "correction" introduces unrealistic magnitudes and even sign reversals for the extracted values of $\xi_{DL}^j$, while neglecting the "correction" for the PMA samples consistently gives results that are in close accord with the HHV results from in-plane magnetic anisotropy samples with the same material components and the similar resistivities and thicknesses and with results from measurements that do not involve the planar Hall effect (such as optical Sagnac interferometry [27] and switching of in-plane spin-orbit torque magnetic tunnel junctions [4]).

### V. Practical impacts

Finally, such [Pt/Co/W]$_n$ superlattices are technologically compelling. As compared in Fig. 3c, the [Pt/Co/W]$_n$ has remarkably higher SOT efficiency than other metallic systems, including Pt-X alloys, Ta and W, while keeping a relatively low resistivity. In Fig. 3d, we demonstrate that the [Pt/Co/W]$_n$ superlattices make it possible to switch rather thick 3$d$ ferromagnets with strong perpendicular magnetic anisotropy (≈9.6 Merg/cm$^3$) using a relatively low current density (about 2.0×10$^7$ A/cm$^2$). This is in contrast to the conventional heavy metal/ferromagnet bilayer scheme, for which a Co layer is unlikely to be switchable by interfacial SOT before the blowout of the device by Joule heating when the thickness is several nm, e.g., 9.6 nm, and can hardly maintain PMA when the effective thickness is above 1.5 nm. The ability of the superlattice to effectively switch a large-volume magnetic layer is interesting for spintronic applications where the magnetic layer has to be thick to maintain a PMA and/or thermal stability when the lateral size scales down to tens of nanometers. Since the power scales with $(1/\xi_{DL}^j)^2$, such [Pt/Co/W]$_n$ superlattices should have $n^2$ times lower power consumption to generate a given spin torque strength than the corresponding conventional magnetic bilayer with the same total magnetic volume.

Therefore, such [Pt/Co/W]$_n$ superlattice can be utilized to develop self-torqued magnetic tunnel junctions and superlattice racetracks for memory,[3,4] computing,[5,6,32] and microwave emitter[5] applications that require low power, long endurance, low impedance, and large magnetic volume for enhanced thermal stability[15,16,33,34] or signal output. In addition, the strong anomalous Nernst effect also makes the superlattices highly desirable for applications, such as thermoelectric batteries for energy harvesting or as sensors of the surface temperature[16].

## IV. Conclusion

In summary, we have demonstrated that, in [Pt/Co/W]$_n$ superlattices with strong PMA, both dampinglike and fieldlike SOTs of the [Pt/Co/W]$_n$ superlattices increase linearly with the repeat number $n$ towards infinite, with $\xi_{DL}^j$=2.25 and $\xi_{FL}^j$ =-0.33 at $n$ = 12. We also find that the dampinglike torque is always much greater than the fieldlike torque, regardless of the number of $n$. These results clarify that the SOTs in the [Pt/Co/W]$_n$ superlattice arise from the spin Hall effect and that there is not any indication of any bulk Rashba-type spin splitting contribution, which establishes a unified understanding of the mechanism and the manipulation of the SOTs in the superlattices. Deterministic switching is, for the first time, achieved in the thicker-than-50-nm PMA [Pt/Co/W]$_{12}$ superlattices with a low current density of 2.0×10$^7$ A/cm$^2$. These results establish the [Pt/Co/W]$_n$ superlattice with strong PMA, giant SOTs, low impendence, good integration compatibility, and strong anomalous Nernst effect as a compelling material candidate for ultra-fast, low-power, long-retention nonvolatile spintronic memory and computing technologies as well as thermoelectric batteries for energy harvesting or as sensors of the surface temperature[16].

**Experimental Section**

[Pt (2.5 nm)/Co (0.8 nm)/W (1.5 nm)]$_n$ superlattices (with $n$ = 1, 4, 6, 8, 10, and 12, respectively) are sputter-deposited onto thermally oxidized silicon substrates with the base pressure of 1× 10$^{-9}$ Torr. Each superlattice was preceded by a 1nm Ta seed layer to enhance adhesion with the substrate and protected from oxidation by a top MgO (1.6 nm)/Ta (1.5 nm) bilayer. The Ta seed layer is highly resistive and has a negligible effect on spin current. For electrical measurements, the samples were patterned into 5 × 60 μm$^2$ Hall bars by photolithography and argon ion milling, followed by the deposition of Ti (5 nm)/Pt (150 nm) contacts. No intentional annealing was performed. The structural properties are characterized using cross-sectional high-angle annular dark-field scanning transmission electron microscopy (HAADF STEM). The magnetization of the samples was measured using a superconducting quantum interference device. The spin-orbit torque is measured by the harmonic Hall voltage technique, with a lock-in amplifier sourcing and detecting the voltage. Keithley source-meters of 2400 and 2450 are used for switching measurements.

**Acknowledgments**

This work is supported partly by the National Key Research and Development Program of China 2022YFA1204000), the Beijing National Natural Science Foundation (Z230006), and the National Natural Science Foundation of China (12304155, 12274405).